\documentclass[10pt]{article} 
\pdfoutput=1
\usepackage[preprint]{rlj} 
\usepackage{amssymb}            
\usepackage{mathtools}          
\usepackage{mathrsfs}           
\usepackage{graphicx}           
\usepackage{subcaption}         
\usepackage[space]{grffile}     
\usepackage{url}                
\usepackage{tablefootnote}
\usepackage{float}
\usepackage{makecell}
\usepackage{amsfonts}
\usepackage{algorithm,algcompatible}
\usepackage{multirow}
\usepackage{flafter}
\usepackage{enumerate}

\title{Learning Multi-Agent Communication Protocol: Study on Information Entropy Efficiency in MARL}

\setrunningtitle{Learning Multi-Agent Communication Protocol: Study on Information Entropy Efficiency in MARL}

\author{Xinren Zhang\textsuperscript{1}, Zixin Zhong\textsuperscript{1}, Jiadong Yu\textsuperscript{1}}

\emails{xzhang231@connect.hkust-gz.edu.cn, \ \{zixinzhong, jiadongyu\}@hkust-gz.edu.cn,}

\affiliations{
$^{1}$\textbf{The Hong Kong University of Science and Technology (Guangzhou), Guangzhou, China}\\
}

\contribution{
1. This paper develops a generalized framework for MARL that accommodates multi-round communication scenarios, providing a unified approach to analyze diverse MARL architectures with communication.}
{
     Prior work has proposed architectures with varying communication rounds-ranging from one-round to multi-round communication-yet lacks a unified structure to systematically characterize communication patterns in MARL~\citep{singh2018learning, niu2021multi}.
    }
\contribution{
2. Our work proposes the IEI that quantifies the relationship between message entropy and task performance, enabling systematic evaluation of how effectively agents minimize information exchange while maintaining task performance.}
{
    Existing work predominantly focuses on task performance without measuring communication efficiency, often achieving performance gains at the cost of increased communication overhead~\citep{sukhbaatar2016learning}.
    }
\contribution{
3. Our work proposes adjustments to policy update mechanisms that incorporate IEI into training loss functions, allowing agents to achieve equivalent or superior task performance while communicating with more compact information.}
{
    Existing approaches predominantly optimize policies based solely on task-related rewards, without incorporating communication efficiency as an explicit regularization term in the training loss function~\citep{li2025assemble}.
    }    

\summary{Multi-Agent Systems (MAS) have emerged as a fundamental paradigm for distributed problem-solving, where autonomous agents collaborate to achieve complex objectives. Within this framework, Multi-Agent Reinforcement Learning (MARL) with communication has demonstrated remarkable success in cooperative tasks. However, existing approaches predominantly pursue performance gains through increasingly complex architectures and expanding communication overhead, lacking principled metrics to evaluate the efficiency of information exchange. In this paper, we focus on enabling agents to learn efficient multi-agent communication protocols that balance performance and information compactness. We propose the Information Entropy Efficiency Index (IEI), a novel metric that quantifies the ratio between message entropy and task performance in learned communication protocols. A lower IEI indicates more compact and efficient message representations. By incorporating IEI into training loss functions, we encourage agents to develop communication protocols that achieve high performance with reduced communication overhead. Extensive experiments across diverse MARL algorithms demonstrate that our approach achieves equivalent or superior task performance compared to baseline methods while improving communication efficiency. These findings challenge the prevailing assumption that performance improvements require complex architectures or increased communication overhead and highlight the potential of improving both task success and communication efficiency to enable scalable MAS.
}

\begin{document}
\maketitle  
\vspace{-.8cm}
\begin{abstract}
\vspace{-.1cm}
Multi-Agent Systems (MAS) have emerged as a fundamental paradigm for distributed problem-solving, where autonomous agents collaborate to achieve complex objectives. Within this framework, Multi-Agent Reinforcement Learning (MARL) with communication has demonstrated remarkable success in cooperative tasks. However, existing approaches predominantly pursue performance gains through increasingly complex architectures and expanding communication overhead, lacking principled metrics to evaluate the efficiency of information exchange. In this paper, we focus on enabling agents to learn efficient multi-agent communication protocols that balance performance and information compactness. We propose the Information Entropy Efficiency Index (IEI), a novel metric that quantifies the ratio between message entropy and task performance in learned communication protocols. A lower IEI indicates more compact and efficient message representations. By incorporating IEI into training loss functions, we encourage agents to develop communication protocols that achieve high performance with improved communication efficiency. Extensive experiments across diverse MARL algorithms demonstrate that our approach achieves equivalent or superior task performance compared to baseline methods while improving communication efficiency. These findings challenge the prevailing assumption that performance improvements require complex architectures or increased communication overhead and highlight the potential of improving both task success and communication efficiency to enable scalable MAS.
\end{abstract}

\vspace{-.7cm}
\section{Introduction}
\vspace{-.3cm}
\label{sec:intro}
Multi-Agent Systems (MAS) represent a fundamental paradigm in artificial intelligence, where collections of autonomous agents interact within shared environments to achieve individual or collective goals~\citep{wooldridge2009introduction}. These systems excel at modeling complex scenarios that range from robotics and traffic management to financial markets and social dynamics. Within this domain, Multi-Agent Reinforcement Learning (MARL) extends traditional reinforcement learning (RL) to environments with multiple interacting agents, introducing fundamental challenges such as nonstationarity, scalability, and coordination~\citep{ning2024survey}. MARL enables agents to develop adaptive cooperative or competitive behaviors for complex tasks that single-agent approaches cannot handle~\citep{sunehag2017value, rashid2020monotonic, foerster2018counterfactual, son2019qtran}. However, coordination becomes particularly challenging in partially observable environments, where agents have incomplete state information, limiting the effectiveness of learned policies.

To overcome this limitation, multi-agent communication has emerged as a crucial mechanism for enhancing coordination and information sharing. Communication protocols define how agents exchange information and can be broadly categorised into engineered, hybrid, and learned approaches. Engineered protocols utilize formal specification languages such as session types and trace expressions, alongside middleware systems like conversation managers that encapsulate structured communication flows~\citep{chopra2020evaluation}. However, the rigidity of predefined protocols limits their adaptability to dynamic environments. Hybrid protocols attempt to bridge this gap through structured agent context protocols and multi-view message certification mechanisms, yet remain constrained by their reliance on predetermined frameworks~\citep{habiba2025revisiting}.

In contrast, learned communication protocols empower agents to discover adaptive communication strategies autonomously. Early frameworks like RIAL and DIAL~\citep{foerster2016learning} established discrete and continuous messaging channels learned end-to-end, while CommNet~\citep{sukhbaatar2016learning} enabled broadcast communication through averaged vector messages. These evolved into sophisticated communication protocols: IC3Net~\citep{singh2018learning} introduced selective communication via gating mechanisms, MAGIC~\citep{niu2021multi} implemented multi-round communication with dynamic scheduling, and attention-based approaches like TarMAC~\citep{das2019tarmac} and G2ANet~\citep{liu2020multi} developed targeted messaging through signature-based attention mechanisms and multiple communication rounds for selective information exchange.

However, existing learned communication protocols predominantly improve task performance while overlooking communication efficiency considerations. Current research implicitly assumes that superior outcomes necessitate more sophisticated network architectures or increased communication overhead, failing to establish whether communication quantity and complexity are the sole determinants of performance limitations. The absence of standardized efficiency metrics prevents systematic comparison of algorithms' communicative effectiveness. Moreover, resource-constrained environments where deploying communication-intensive systems is impractical~\citep{10342771,chafii2023emergent} suggest that current research may overlook more efficient approaches.

To address this critical gap, we propose the Information Entropy Efficiency Index (IEI), a novel metric quantifying the ratio between average message entropy and task performance. Lower IEI values indicate more compact information utilization, enabling systematic evaluation of how effectively agents encode meaningful information under fixed communication constraints. By incorporating IEI directly into loss functions for policy update, we introduce communication protocols that maximize information utility rather than quantity, transforming communication efficiency from a constraint into an explicit part of the optimization objective that enhances the task performance of MAS through more effective information exchange.
The contributions of this paper can be summarized as follows:
\begin{itemize}
\item \textbf{Develop} a generalized framework for MARL that accommodates multi-round communication scenarios, providing a unified approach to analyze diverse MARL architectures with communication.
\item \textbf{Propose} the IEI that quantifies the relationship between message entropy and task performance, enabling systematic evaluation of how effectively agents minimize information exchange while maintaining task performance.
\item \textbf{Propose} adjustments to policy update mechanisms that incorporate IEI into training loss functions, allowing agents to achieve equivalent or superior task performance while communicating with more compact information.
\end{itemize}

The paper is organized as follows: Section~\ref{Sec2:Related Works} shows the related works of MARL considering communication. Section~\ref{sec:MODEL} introduces our generalized MARL model of learning communication protocol. Section~\ref{sec:IEI} proposes the IEI for evaluating communication efficiency and presents comparative experimental results across different MARL baseline algorithms. Section~\ref{sec:loss} demonstrates how incorporating IEI into loss functions during policy update enhances both communication efficiency and task performance. Finally, Section~\ref{sec:conclusion} summarizes our contributions and findings.
\vspace{-.3cm}
\section{Related Work}
\label{Sec2:Related Works}
\vspace{-.3cm}
MARL with learned communication protocols enables agents to autonomously develop coordination strategies through experience. These protocols comprise two key components: communication topology (when and with whom to communicate) and message content (what information to transmit). We review existing approaches by examining how these components are learned, with particular focus on attention mechanisms as a prevalent technical tool for enhancing protocol design.
\vspace{-.3cm}
\subsection{Communication Protocol Learning}
\vspace{-.2cm}
Existing learned communication protocols can be categorized based on which protocol aspects they address: topology control, message content design, or their joint optimization.

\textbf{Topology Control.} Early work by Foerster et al.~\citep{foerster2016learning} established learnable communication topology through RIAL and DIAL, which enable agents to condition decisions on cross-timestep message exchanges. Recognizing that permanent communication links may be inefficient, Singh et al.~\citep{singh2018learning} developed IC3Net with a gating mechanism that learns when communication is necessary. SchedNet~\citep{kim2019learning} further advanced selective communication through a weight-based scheduler that dynamically determines broadcasting agents at each timestep. More recently, MAGIC~\citep{niu2021multi} modeled communication as a learnable directed graph, using a scheduler to produce adjacency matrices for targeted multi-round message passing.

\textbf{Message Content Design.} The predominant paradigm focuses on hidden state sharing, where agents broadcast internal representations as communication messages. CommNet~\citep{sukhbaatar2016learning} exemplified this approach, enabling fully cooperative agents to learn communication by averaging hidden states across agents. HetNet~\citep{seraj2022learning} extended this by generating class-specific hidden states that are transformed and optionally binarized for bandwidth efficiency, with receiving agents using class-adapted attention coefficients for message aggregation.

\textbf{Joint Learning.} Recent approaches recognize the interdependence of topology and content. TarMAC~\citep{das2019tarmac} implemented targeted communication where agents learn both message content and recipient selection through signature-value pairs and multi-round attention-based addressing. G2ANet~\citep{liu2020multi} employed a two-stage design combining hard attention for topology selection and soft attention for message weighting, enabling efficient communication through graph neural network structures.
\vspace{-.2cm}
\subsection{Attention-Enhanced Communication}
\vspace{-.2cm}

Attention mechanisms have emerged as a key technical approach for improving both topology control and message aggregation, overcoming the inefficiency of indiscriminate broadcasting through selective information processing.

For topology selection, attention enables dynamic determination of communication partners. ATOC~\citep{jiang2018learning} used attention to decide if agents should communicate within their observable field, while MAGIC~\citep{niu2021multi} employed attention-based scheduling to produce adjacency matrices specifying directed communication links. G2ANet~\citep{liu2020multi} implemented hard attention with binary gating to identify relevant interactions, establishing sparse connectivity patterns that adapt to task requirements.

For message aggregation, attention mechanisms enable intelligent processing of received messages. TarMAC~\citep{das2019tarmac} used signature-based soft attention where agents compute attention weights through query-signature dot products, enabling targeted information extraction. G2ANet~\citep{liu2020multi} employed soft attention to weight agent contributions through weighted sums of hidden states, while MAGIC~\citep{niu2021multi} used attention-based message processors to integrate messages for decision making. This attention-weighted aggregation allows agents to prioritize relevant information while filtering less pertinent signals.

Despite these advances, evaluation of learned communication frameworks often lacks practical insights for real-world deployment where resource constraints are critical. Additionally, the trade-offs between one-round and multi-round communication remain underexplored, highlighting the need for systematic evaluation considering both communication efficiency and task performance in resource-constrained environments.

\vspace{-.3cm}
\section{A Developed General Framework for Learning Communication Protocols in MARL}\label{sec:MODEL}
\vspace{-.3cm}
\begin{figure}[t]
  \centering
  \includegraphics[width=1\textwidth]{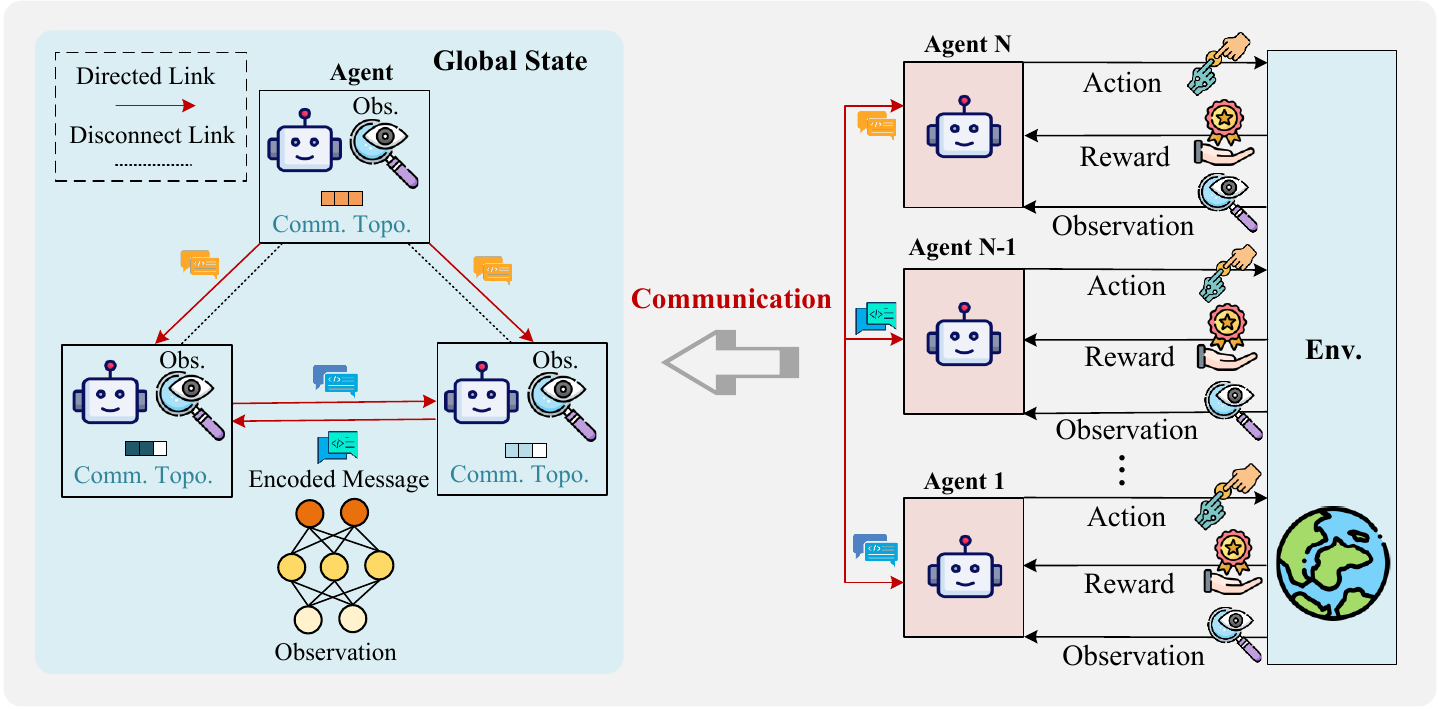}
  \caption{Illustration of the Communications in MARL.}
  \label{fig1}
  \vspace{-.3cm}
\end{figure}
Communication serves as a fundamental mechanism for enhancing coordination and information sharing in MAS. Fig.~\ref{fig1} illustrates the communication process in MARL, indicating how agents collaborate to accomplish tasks through message exchange. The system operates with a communication topology that determines which agents can exchange information. Under a shared global state, each agent receives only partial observations of the environment, which they process to generate encoded messages. The communication topology then governs message routing, determining which specific agents receive each message. Upon receiving both local observations and messages from neighbors, each agent decides actions accordingly, with the environment subsequently providing rewards based on collective actions.

By expanding agents' effective observation space beyond local views, this communication framework enables agents to learn what information to broadcast and to whom, improving situational awareness and reducing uncertainty for coordinated behavior. This learned mechanism proves particularly valuable in scenarios with heterogeneous agents or non-stationary objectives where predefined rules fail to generalize. To enable such coordination in complex environments, we develop a generalized CTDE-based multi-round communication framework. Agents iteratively exchange messages during centralized training to refine joint decision-making policies, while the learned communication protocols enable fully decentralized execution at deployment, maintaining autonomy and scalability.
\vspace{-.2cm}
\subsection{Centralized Training Stage}
\vspace{-.2cm}
During the training phase, global information is leveraged to facilitate more effective learning while ensuring the execution phase remains fully decentralized. The centralized training framework consists of the following integrated components:

\textbf{Observation Processing.}
Each agent $i \in [0,N-1]$ receives its local observation $o_i^t$ from the environment with global state $s_t$. During the centralized training phase, these observations are processed collectively. The collective observations $\mathbf{o}^t = [o_0^t, o_1^t, \ldots, o_{N-1}^t]$ are transformed through a shared observation encoder into initial hidden states:
\begin{equation}
    \mathbf{h}^{t^{(0)}} = f_{\boldsymbol{\theta}_\text{obs}}^o(\mathbf{o}^t),
\end{equation}
where $f_{\boldsymbol{\theta}_\text{obs}}^o(\cdot)$ is the observation encoding function with learnable parameters $\boldsymbol{\theta}_\text{obs}$, typically implemented as a neural network that transforms raw observations into feature representations, and $\mathbf{h}^{t^{(0)}} = [h_0^{t^{(0)}}, h_1^{t^{(0)}}, \ldots, h_{N-1}^{t^{(0)}}]$ represents the collection of all agents' initial hidden states.

\textbf{Multi-Round Communication Protocol.}
The training process incorporates $L$ iterations of information exchange among agents before action selection. For each communication round $l \in \{1,2,\cdots,L\}$, agents engage in the following systematic steps:

\begin{itemize}
\item Step 1: Message Encoding. Each agent encodes its current hidden state from the previous round into an initial message vector:
\begin{equation}
    \mathbf{m}^{*t^{(l)}} = f_{\boldsymbol{\theta}_\text{msg}}^m(\mathbf{h}^{t^{(l-1)}}),
\end{equation}
where $f_{\boldsymbol{\theta}_\text{msg}}^m(\cdot)$ is the message encoding function parameterized by $\boldsymbol{\theta}_\text{msg}$, which transforms hidden states into message representations suitable for communication, and $\mathbf{m}^{*t^{(l)}} = [m_0^{*t^{(l)}}, m_1^{*t^{(l)}}, \ldots, m_{N-1}^{*t^{(l)}}]$ represents the collection of all agents' initial messages.
This encoding step can be implemented through various strategies: identity transformation for direct hidden state broadcasting~\citep{sukhbaatar2016learning}, linear projections for dimensionality adaptation, or learned transformations with class-specific encoders~\citep{seraj2022learning} that generate specialized representations for heterogeneous agents.

\item Step 2: Communication Topology Selection. The centralized controller determines the communication graph $G^{t^{(l)}} \in \mathbb{R}^{N \times N}$ based on the latest hidden state:
\begin{equation}
    G^{t^{(l)}} = f_{\boldsymbol{\theta}_\text{topo}}^\text{topo}(\mathbf{h}^{t^{(l-1)}}),
\end{equation}
where $f_{\boldsymbol{\theta}_\text{topo}}^\text{topo}(\cdot)$ is the topology selection function with parameters $\boldsymbol{\theta}_\text{topo}$, which learns to determine optimal communication links based on current hidden states, and $G_{i,j}^{t^{(l)}}=1$ indicates that agent $i$ establishes a communication channel with agent $j$.
The topology function can take multiple forms depending on the desired communication structure. Early approaches~\citep{foerster2016learning} enabled cross-timestep broadcasting through recurrent architectures, while gating-based methods~\citep{singh2018learning} learn binary decisions on whether to activate communication channels. Scheduler-based approaches~\citep{kim2019learning} determine which subset of agents should broadcast at each step, and graph-based methods~\citep{niu2021multi} produce adjacency matrices specifying directed links for targeted message passing. Attention mechanisms~\citep{jiang2018learning} can also be employed to identify relevant communication partners within observable neighborhoods.

\item Step 3: Message Aggregation. Based on the initial message vectors and the communication graph, the center integrates information from connected agents:
\begin{equation}
    \mathbf{m}^{t^{(l)}} = f_{\boldsymbol{\theta}_\text{aggr}}^\text{aggr}(\mathbf{m}^{*t^{(l)}}, G^{t^{(l)}}),
\end{equation}
where $f_{\boldsymbol{\theta}_\text{aggr}}^\text{aggr}(\cdot)$ is the message aggregation function parameterized by $\boldsymbol{\theta}_\text{aggr}$, which combines messages from connected agents according to the communication topology, and $\mathbf{m}^{t^{(l)}} = [m_0^{t^{(l)}}, m_1^{t^{(l)}}, \ldots, m_{N-1}^{t^{(l)}}]$ represents the collection of all agents' aggregated messages.
Aggregation mechanisms vary in sophistication: simple averaging~\citep{sukhbaatar2016learning} treats all messages equally, while attention-based aggregation~\citep{das2019tarmac, liu2020multi, niu2021multi} computes weighted combinations where agents prioritize relevant information through learned attention coefficients. Query-key mechanisms~\citep{das2019tarmac} enable targeted extraction by matching agent queries with message signatures, while hard attention~\citep{liu2020multi} applies binary selection to establish sparse connectivity patterns.

\item Step 4: Hidden State Update. Following message aggregation, each agent updates its hidden state based on its previous hidden state and the aggregated message information:
\begin{equation}
    \mathbf{h}^{t^{(l)}} = f_{\boldsymbol{\theta}_\text{hsu}}^\text{hsu}(\mathbf{h}^{t^{(l-1)}}, \mathbf{m}^{t^{(l)}}),
\end{equation}
where $f_{\boldsymbol{\theta}_\text{hsu}}^\text{hsu}(\cdot)$ is the hidden state update function with learnable parameters $\boldsymbol{\theta}_\text{hsu}$, typically implemented as a recurrent neural network or transformer layer that integrates previous hidden states with received messages, and $\mathbf{h}^{t^{(l)}} = [h_0^{t^{(l)}}, h_1^{t^{(l)}}, \ldots, h_{N-1}^{t^{(l)}}]$ represents the updated hidden states for all agents.
\end{itemize}

This process repeats for $L$ communication rounds. Although the topology generation parameters $\boldsymbol{\theta}_\text{topo}$ remain fixed throughout all rounds, the communication graph $G^{t^{(l)}}$ evolves dynamically across rounds because it takes as input the continuously updated hidden states $\mathbf{h}^{t^{(l-1)}}$ from the previous round. Consequently, while the topology generation mechanism is shared, the actual communication topologies adapt to the evolving information state of the system.
This multi-round iterative process~\citep{niu2021multi, das2019tarmac} enables progressive information refinement, where agents gradually incorporate knowledge from increasingly informed neighbors, facilitating more sophisticated coordination than single-round protocols.

\textbf{Policy Optimization.}
Upon completion of all $L$ communication rounds, the center utilizes the final hidden states to formulate policies and select actions for all agents:
\begin{equation}
    \mathbf{a}^t = f_{\boldsymbol{\theta}_\pi}^{\pi}(\mathbf{h}^{t^{(L)}}),
\end{equation}
where $f_{\boldsymbol{\theta}_\pi}^{\pi}(\cdot)$ is the policy function parameterized by $\boldsymbol{\theta}_\pi$, which maps the final hidden states to action distributions or deterministic actions, and $\mathbf{a}^t = [a_0^t, a_1^t, \ldots, a_{N-1}^t]$ represents the joint action of all agents.

During the centralized training phase, we improve both the individual policies and the communication mechanisms simultaneously. This is achieved through a centralized critic that has access to global state information:
\begin{equation}
    Q(s_t, \mathbf{a}^t) = f_{\boldsymbol{\theta}_Q}^{Q}(s_t, \mathbf{a}^t),
\end{equation}
where $f_{\boldsymbol{\theta}_Q}^{Q}(\cdot)$ is the centralized critic function with parameters $\boldsymbol{\theta}_Q$, which estimates the state-action value function using global state and joint action information.

\textbf{Training Loss.}
The training process optimizes the following comprehensive objective function
\begin{equation}
    \mathcal{L}_t = l_{\mathbf{a}^t} + w_q l_{Q_t},
    \label{eq:loss}
\end{equation}
where $l_{\mathbf{a}^t}$ is the policy loss and $l_{Q_t}$ is the critic loss, with $w_q$ being the weighting coefficient that balances the two loss components.
The network parameters are updated using gradient descent:
\begin{equation}
    \boldsymbol{\theta} \leftarrow \boldsymbol{\theta} - \eta \nabla_{\boldsymbol{\theta}} \mathcal{L}_t,
\end{equation}
where $\boldsymbol{\theta} = \{\boldsymbol{\theta}_\text{obs}, \boldsymbol{\theta}_\text{msg},
\boldsymbol{\theta}_\text{topo},
\boldsymbol{\theta}_\text{aggr}, \boldsymbol{\theta}_\text{hsu}, \boldsymbol{\theta}_\pi, \boldsymbol{\theta}_Q\}$ represents the collective parameters of all network components, and $\eta$ is the learning rate.
The end-to-end differentiability of this framework~\citep{foerster2016learning, sukhbaatar2016learning} enables gradient flow through communication channels during training, allowing joint optimization of both coordination strategies and communication protocols.
\vspace{-.2cm}
\subsection{Decentralized Execution Stage}
\vspace{-.2cm}
During the decentralized execution stage, each agent $i\in [0,N-1]$ operates independently using the parameters $\boldsymbol{\theta}$ learned from the centralized training phase. These parameters remain fixed during execution, with no further updates occurring as agents interact with the environment.
Each agent receives its local observation, processes it through the trained neural networks, engages in the multi-round communication protocol, and makes decisions based solely on its own observations and the messages received from other agents, without access to global state information or centralized control.

\vspace{-.2cm}
\subsection{Experiment Evaluation of Different Baselines within Different Rounds}
\begin{figure}[h]
  \centering
  \includegraphics[width=1\textwidth]{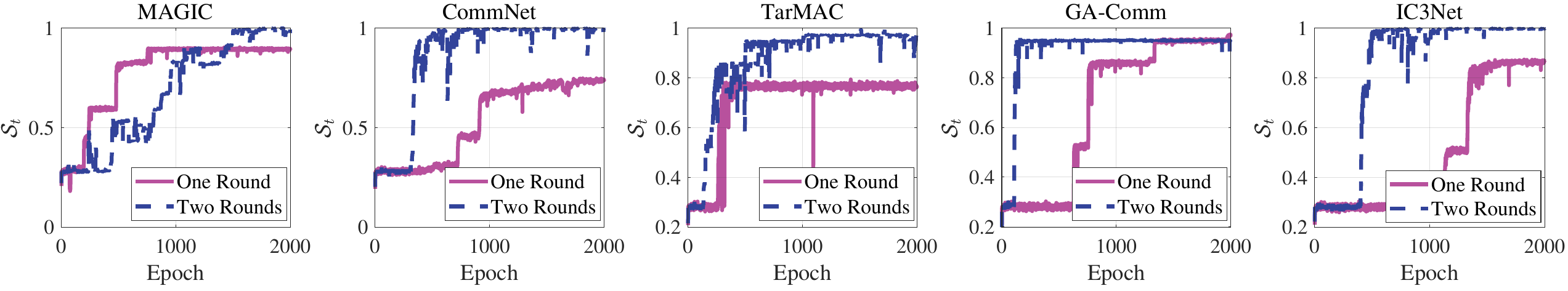}
  \caption{Success Rate Comparison Across Baselines: One-Round ($L=1$) vs. Two-Round ($L=2$) Communication.}
  \label{fig2}
  \vspace{-.8cm}
\end{figure}

\vspace{-.2cm}
\begin{table}[h]
\centering
\caption{Environment Parameters of TJ}
\label{table1}
\begin{tabular}{l  l  l  l }
\hline
\textbf{Parameters} & \textbf{Value}&\textbf{Parameters} & \textbf{Value} \\ \hline
Number of Agents  & $5$ & Grid Size &$7\times7$\\
Steps per Episode &20 & Number of Epochs & $2000$ \\
\makecell[l]{Number of Batches per Epoch}  & $10$  &\makecell[l]{Number of Episodes per Batch}  & $500$ \\
Vision Range &1 &Arriving Prob. of Cars & $0.3$ \\
\hline
\end{tabular}
\vspace{-.2cm}
\end{table}
Based on the developed framework, we conduct a comprehensive study of five classical multi-agent communication algorithms -- MAGIC~\citep{niu2021multi}, CommNet~\citep{sukhbaatar2016learning}, TarMAC~\citep{das2019tarmac}, GA-Comm~\citep{liu2020multi}, and IC3Net~\citep{singh2018learning} -- within the Traffic Junction (TJ) environment~\citep{singh2018learning}. We compare the success rate of different baselines within one-round and two-round communication. The parameter values of the TJ environment can be found in Table~\ref{table1}. The comparison results are shown in Fig. \ref{fig2}.

As shown in Fig. \ref{fig2}, we observe a fundamental trade-off between communication efficiency and task performance in MAS. Increasing communication rounds generally improves final success rates across most baselines, though at the cost of increased communication costs and computational complexity.

When transitioning from one-round to two-round communication, MAGIC, CommNet, TarMAC, and IC3Net all exhibit substantial improvements in asymptotic success rates, with particularly pronounced gains in CommNet and IC3Net. These improvements suggest that additional communication rounds enable agents to refine coordination strategies through iterative information exchange, facilitating more sophisticated reasoning about joint actions in complex cooperative tasks.

GA-Comm presents a notable exception: while its final success rate under two-round communication remains comparable to one-round performance, it achieves convergence by approximately epoch 200-substantially earlier than the one-round variant. This accelerated convergence indicates that multi-round communication can enhance sample efficiency even when asymptotic performance remains similar.

These observations confirm that increasing communication rounds enhances both coordination quality and learning efficiency. However, this comes with substantial practical constraints. Multi-round protocols impose multiplicative increases in bandwidth requirements, latency, and energy consumption-factors that become prohibitive as agent populations scale. In resource-constrained environments such as robotic swarms or wireless sensor networks, the cumulative overhead may exceed available resources or violate real-time constraints.

This tension between coordination performance and communication efficiency motivates our central challenge: developing protocols that achieve high coordination quality while maintaining improved communication efficiency. Yet, a fundamental question remains: how do we systematically evaluate whether an algorithm achieves this balance? Existing research predominantly assesses algorithms through task performance metrics alone, such as success rate or cumulative reward, without accounting for the communication resources required to achieve these outcomes. This evaluation paradigm obscures a critical dimension of algorithm efficiency-whether performance gains justify their communication costs, particularly as systems scale to larger agent populations.
\vspace{-.4cm}
\section{Proposal of Information Entropy Efficiency Index}
\label{sec:IEI}
\vspace{-.4cm}
Effective multi-agent coordination fundamentally requires balancing two competing objectives: achieving high task performance while minimizing communication resource consumption. Current research primarily evaluates algorithms based solely on task performance metrics. To address this evaluation gap, we propose the IEI, a metric that quantifies the message entropy required per unit of task success. By measuring the relationship between message entropy and success rate, IEI enables systematic assessment of how efficiently agents encode task-relevant information, with lower values indicating more compact communication strategies. This metric reveals not only coordination quality but also the efficiency of information encoding to achieve that coordination.
\vspace{-.4cm}
\subsection{Information Entropy Efficiency Index (IEI)}
\vspace{-.3cm}
This metric quantifies the amount of entropy required per unit of success:
\begin{equation}
    \Phi_{{\text{IEI}}_t} = \frac{H_t}{\mathscr{S}_t},
\end{equation}
where $\mathscr{S}_t$ represents the success rate at epoch $t$, and $H_t$ is the average message entropy across all agents and communication rounds in epoch $t$:
\begin{equation}
    H_t = \frac{1}{L} \sum_{l=1}^{L} \left( \frac{1}{N} \sum_{i=0}^{N-1} H(M_i^{t^{(l)}}) \right).
\end{equation}
The entropy of each agent's communication message is calculated as:
\begin{equation}
    H(M_i^{t^{(l)}}) = -\sum_{k} p(m_{i,k}^{t^{(l)}}) \log_2 p(m_{i,k}^{t^{(l)}}),
\end{equation}
where $p(m_{i,k}^{t^{(l)}})$ represents the normalized probability distribution of agent $i$'s message components during the $l^{\text{th}}$ communication round. \footnote{We compute message entropy by treating each dimension of the continuous message vector as a discrete event, where the normalized absolute values serve as the probability distribution (i.e., $p(m_{i,k}^{(l)}) = |m_{i,k}^{(l)}| / \sum_k{|m_{i,k}^{(l)}|}$). This approach quantifies the concentration of information across message dimensions without requiring explicit discretization.}

A lower $\Phi_{{\text{IEI}}_t}$ indicates that agents achieve success while exchanging information with lower entropy, suggesting more efficient encoding of task-relevant information.
\vspace{-.3cm}
\subsection{Experiment Evaluation of IEI}
\vspace{-.3cm}
\begin{figure}[h]
    \centering
    \includegraphics[width=.8\linewidth]{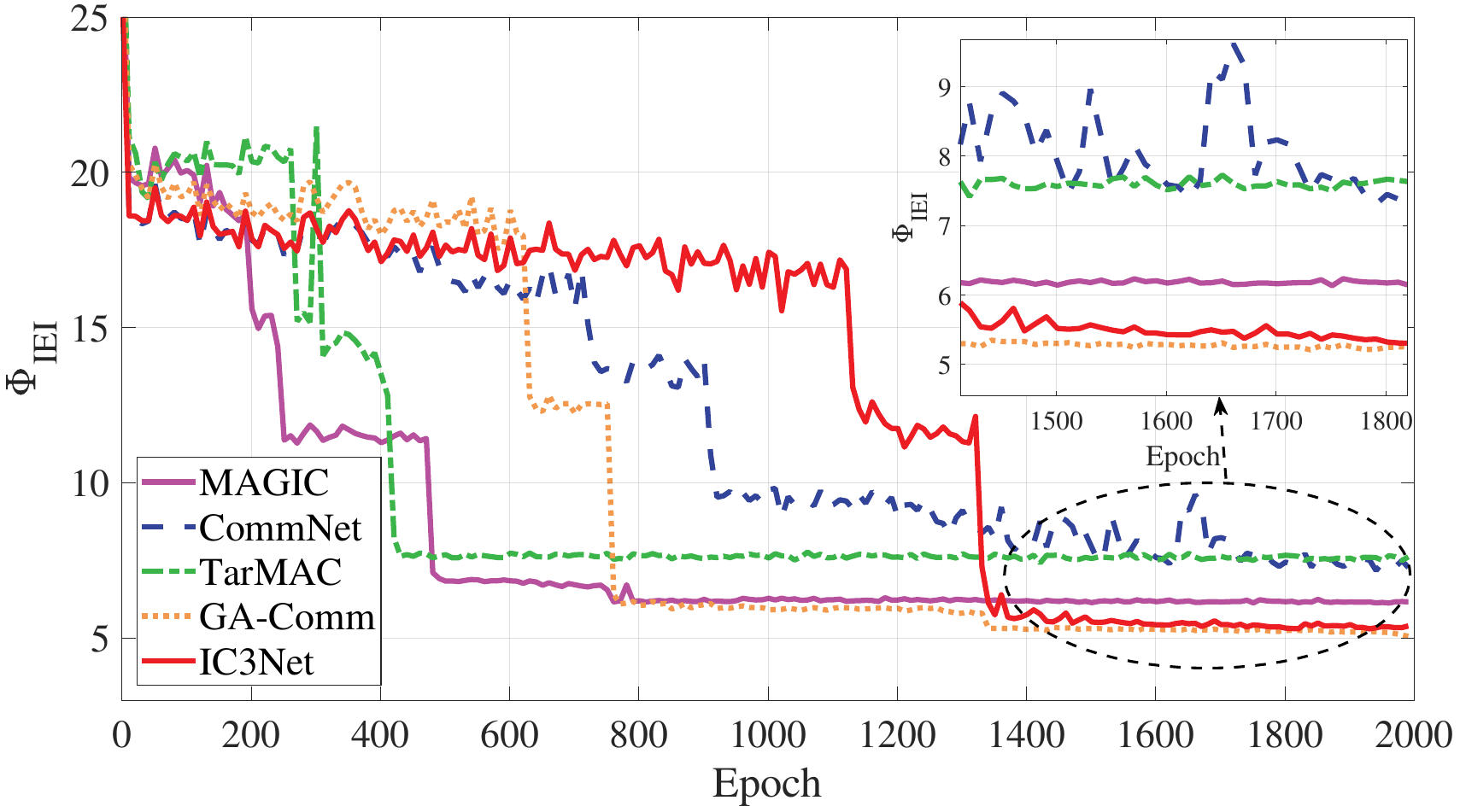}
\caption{Comparison of $\Phi_{{\text{IEI}}}$ for Different Algorithms in the TJ Environment with $L=1$.}  
    \label{fig:TJ_one_round_Index2_mean}
\end{figure}
Based on the proposed metric, we conduct a comprehensive study of five classical multi-agent communication algorithms within the TJ environment. Our experiments apply $\Phi_{{\text{IEI}}}$ to systematically evaluate the effectiveness of communication in conveying task-critical information. While our framework supports $L$-round communication, we focus our analysis on the one-round scenario ($L=1$) to evaluate communication efficiency under minimal communication overhead conditions. This choice allows us to examine how algorithms perform when communication resources are most constrained, providing insights into their fundamental efficiency characteristics and learning dynamics. 
\begin{figure}[t]
    \centering
    \includegraphics[width=1\linewidth]{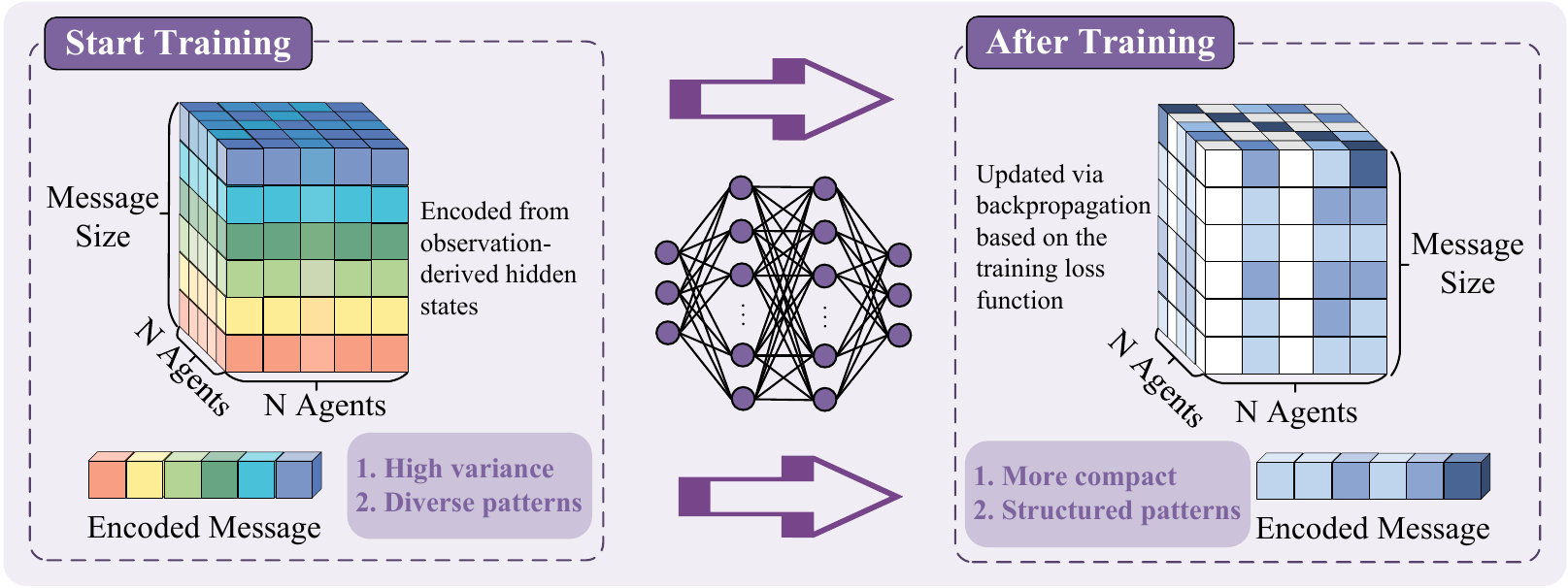}
\caption{Illustration of the Proposed IEI Changes During Training.}  
    \label{fig:IEI_Illustration}
    \vspace{-.3cm}
\end{figure}

Fig. \ref{fig:TJ_one_round_Index2_mean} compares the $\Phi_{{\text{IEI}}}$ value of five communication algorithms in one-round scenarios, where lower values indicate more focused communication.
All algorithms begin with high entropy levels (20-23) but follow distinct reduction patterns. TarMAC achieves the earliest significant entropy reduction around epoch 300. MAGIC displays step-wise reductions, eventually stabilizing around 6. CommNet shows more gradual decreases. Notably, GA-Comm and IC3Net ultimately achieve the lowest terminal entropy values (approximately 5) by the end of training, though IC3Net maintains high entropy longest before converging after epoch 1300. These empirical findings show that the IEI value declines as training epochs progress, inspiring a method to enhance agents' communication efficiency.

Fig. \ref{fig:IEI_Illustration} illustrates the evolution of the proposed IEI during training based on observations from Fig. \ref{fig:TJ_one_round_Index2_mean}. As shown, the encoded messages transmitted by agents undergo a significant transformation: initially exhibiting high variance and diverse patterns across the feature space, they gradually converge toward a more compact and structured representation as training progresses. This convergence demonstrates that agents learn to encode information more efficiently, transitioning from scattered, heterogeneous messages to organized, homogeneous communication patterns.

\vspace{-.3cm}
\section{Learning Protocol with Communication Efficiency}
\label{sec:loss}
\vspace{-.3cm}
IEI observations reveal that encoded messages progressively converge toward more compact and structured representations throughout training, indicating emergent communication efficiency. Building on this insight, we propose an enhanced optimization framework that explicitly incorporates this efficiency metric into the training loss function, accelerating communication efficiency development while maintaining or enhancing overall task performance.
\vspace{-.2cm}
\subsection{Communication Efficiency-Augmented Loss Function}
\vspace{-.2cm}
We extend the original objective in Eq.~(\ref{eq:loss}) with an additional term that explicitly incentivizes communication efficiency:
\begin{equation}
    \mathcal{L}_t = l_{\mathbf{a}^t} + w_ql_{{Q_t}} + w_{{\text{IEI}}_t}\Phi_{{\text{IEI}}_t},
\end{equation}
where
$\Phi_{{\text{IEI}}_t}$ corresponds to the IEI,
$w_q$ and $w_{{\text{IEI}}_t}$ are hyperparameters that balance the different objectives.
By incorporating IEI directly into the loss function, we enhance the learning process toward more efficient communication protocols.

\begin{algorithm}[h] \label{Algo1}
\caption{Communication Efficiency Regularization}
\begin{algorithmic}[1]
\STATE {\bfseries Input:} Original loss $\mathcal{L}_t = l_{\mathbf{a}^t} + w_ql_{{Q_t}}$, average message entropy $H_t$ across all agents and communication rounds in epoch $t$, success rate $\mathscr{S}_t$.
\STATE Set minimum success rate threshold $\mathcal{T}$, scaling factor $\beta$, regularization target ratio $\alpha$, minimum weight $\lambda_\text{min}$, maximum weight $\lambda_\text{max}$, and small constant $\epsilon$.
\IF{$\mathscr{S}_t < \mathcal{T}$}
    \STATE $\mathscr{S}_t \leftarrow \mathcal{T}$
\ENDIF

\STATE Compute IEI in epoch $t$: $\Phi_{{\text{IEI}}_t} = H_t \cdot (1.0 - \beta \cdot \mathscr{S}_t)$ and 
dynamic weight: $w_{{\text{IEI}}_t} = \alpha \cdot \mathcal{L}_t / (\Phi_{{\text{IEI}}_t} + \epsilon)$.
Constrain weight range: $w_{{\text{IEI}}_t} = \max(\lambda_\text{min}, \min(\lambda_\text{max}, w_{{\text{IEI}}_t}))$.
\STATE Update total loss: $\mathcal{L}_t = \mathcal{L}_t + w_{{\text{IEI}}_t} \cdot \Phi_{{\text{IEI}}_t}$.
\STATE {\bfseries Output:} Updated total loss $\mathcal{L}_t$.
\end{algorithmic}
\noindent{\textbf{Note:} The formulation $\Phi_{{\text{IEI}}_t} = H_t \cdot (1.0 - \beta \cdot \mathscr{S}_t)$ provides a smoother regularization mechanism compared to the direct division by success rate ($H_t/\mathscr{S}_t$). This approach reduces sensitivity of $\Phi_{{\text{IEI}}_t}$ to success rate fluctuations -- applying milder regularization when the success rate approaches 1, while preventing excessive regularization intensity when success rates are low.}
\end{algorithm}

To address the challenge of balancing task performance with communication efficiency, we implement a dynamic regularization weight adjustment mechanism that ensures communication efficiency is appropriately emphasized without compromising primary task objectives. The dynamic regularization weight adjustment mechanism is shown in Algorithm 1. This dynamic weighting strategy automatically adjusts regularization intensity based on current task performance while maintaining proportionality between the primary learning objective and efficiency regularization terms. This mechanism enables a natural progression in training priorities. During early stages when task performance is poor, regularization pressure is automatically reduced, allowing agents to focus primarily on task completion. As performance stabilizes and improves, the system gradually increases emphasis on communication efficiency. This adaptive approach achieves a well-calibrated balance between task performance and communication efficiency throughout the learning process.

In other words, when agents achieve higher success rates, the system can afford to place greater emphasis on communication efficiency. Conversely, when performance deteriorates, the regularization pressure is automatically reduced, allowing agents to prioritize task completion. This adaptive approach prevents scenarios where excessive regularization might impede learning progress during challenging phases of training. 
\vspace{-.2cm}
\subsection{Implementation Details}
\vspace{-.2cm}
Afterward, the numerical experiments are conducted in the TJ environment with the same parameter setting as in Section~\ref{sec:MODEL} to evaluate the performance of MARL algorithms before and after the aforementioned changes of the loss function. Our experiments were implemented in Python 3.10.16 using PyTorch. All training and evaluation procedures were executed on an Intel(R) Core(TM) i7-14650HX processor with an NVIDIA GeForce RTX 4060Ti Laptop GPU. The key experimental parameters are detailed in Table~\ref{table2}.
\begin{table}[ht]
\centering
\caption{Simulation parameters}
\label{table2}
\renewcommand{\arraystretch}{1.3}  
\begin{tabular}{l  l  l  l }
\hline
\textbf{Parameters} & \textbf{Value}&\textbf{Parameters} & \textbf{Value} \\ \hline
$\epsilon$  & $10^{-10}$ & $\mathcal{T}$ &0.05\\
$\beta$ &0.5 & $\alpha$ & $0.01$ \\
$\lambda_{\text{min}}$  & $10^{-5}$  &$\lambda_{\text{max}}$  & $5 \times 10^{-3}$ \\
\hline
\end{tabular}
\vspace{-.4cm}
\end{table}
\vspace{-.2cm}
\subsection{Experiment Results}
\begin{figure}[t]
    \centering
    \includegraphics[width=1\linewidth]{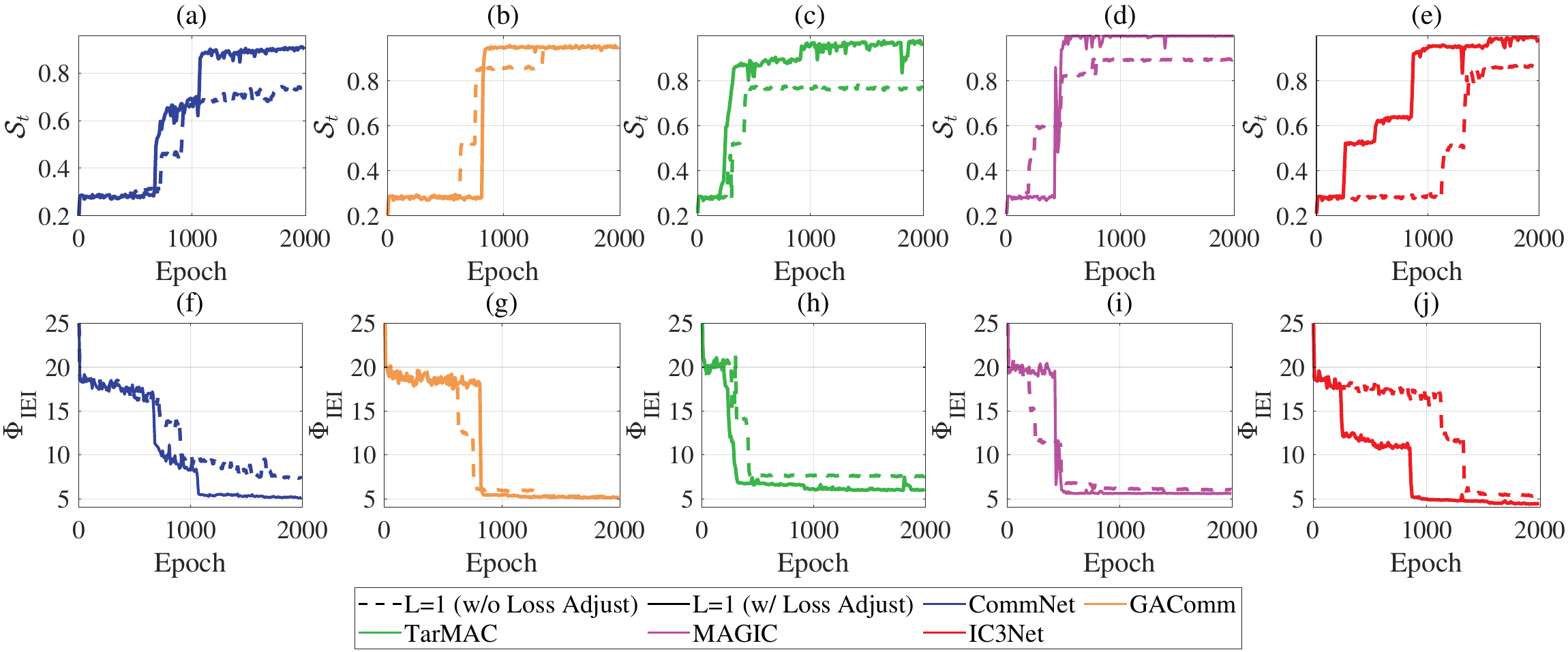}
\caption{Comparison of Success Rate and $\Phi_{{\text{IEI}}}$ for Different Algorithms in the TJ Environment with and without Loss Adjustment (One-Round Communication Scenario).}  
    \label{fig:Loss_Adjust}
    \vspace{-.4cm}
\end{figure}

\vspace{-.2cm}
Fig.~\ref{fig:Loss_Adjust} presents the comparative success rates ($\mathscr{S}_t$) and $\Phi_{{\text{IEI}}}$ of various multi-agent communication algorithms in the TJ 
environment, contrasting performance with and without loss adjustment
mechanisms during one-round communication for each MARL algorithm. The experimental results reveal heterogeneous algorithmic responses to loss adjustment, with each algorithm exhibiting distinct optimization trajectories that provide comprehensive insights into the mechanism's effectiveness. 

In Fig.~\ref{fig:Loss_Adjust} (a) (f), CommNet demonstrates a unique pattern where loss adjustment delays convergence from 971 to 1091 epochs while substantially enhancing final performance from 0.74 to 0.90, representing a 22\% improvement that prioritizes solution quality over convergence speed, accompanied by a 27\% reduction in final IEI values from 7.5 to 5.5. In Fig.~\ref{fig:Loss_Adjust} (b) (g), GAComm achieves the convergence acceleration from 1371 to 841 epochs (1.6 times faster) while maintaining consistent high performance around 0.94 and identical final IEI efficiency of 5.21, though with notably faster stabilization patterns. 

As shown in Fig.~\ref{fig:Loss_Adjust} (c) (h), TarMAC exhibits remarkable transformation with 2.3 times faster convergence acceleration and performance enhancement from 0.76 to 0.96, coupled with IEI improvements from 7.71 to 6.86 achieved through faster convergence from 471 to 321 epochs. MAGIC (Fig.~\ref{fig:Loss_Adjust} (d) (i)) demonstrates exceptional optimization with convergence improvement from 800 to 591 epochs and dramatic performance gains from 0.893 to more than 0.999, while simultaneously reducing IEI from 6.71 to 5.59, representing a 17\% efficiency enhancement. IC3Net (Fig.~\ref{fig:Loss_Adjust} (e) (j)) achieves balanced improvements with 1.5 times faster convergence and substantial performance gains from 0.85 to more than 0.98, accompanied by accelerated IEI convergence from 1391 to 881 epochs and modest efficiency improvements from 5.67 to 5.21. 

These observations indicate that the aforementioned adjustment of loss functions differentially benefits algorithms with their inherent architectural characteristics. Rather than solely accelerating existing learning trajectories, it enables a principled balance between task performance and communication efficiency. The consistent IEI improvements across all algorithms, coupled with varying convergence patterns, demonstrate that communication efficiency and task effectiveness can be simultaneously optimized.
\begin{figure}[t]
    \centering
    \includegraphics[width=.8\linewidth]{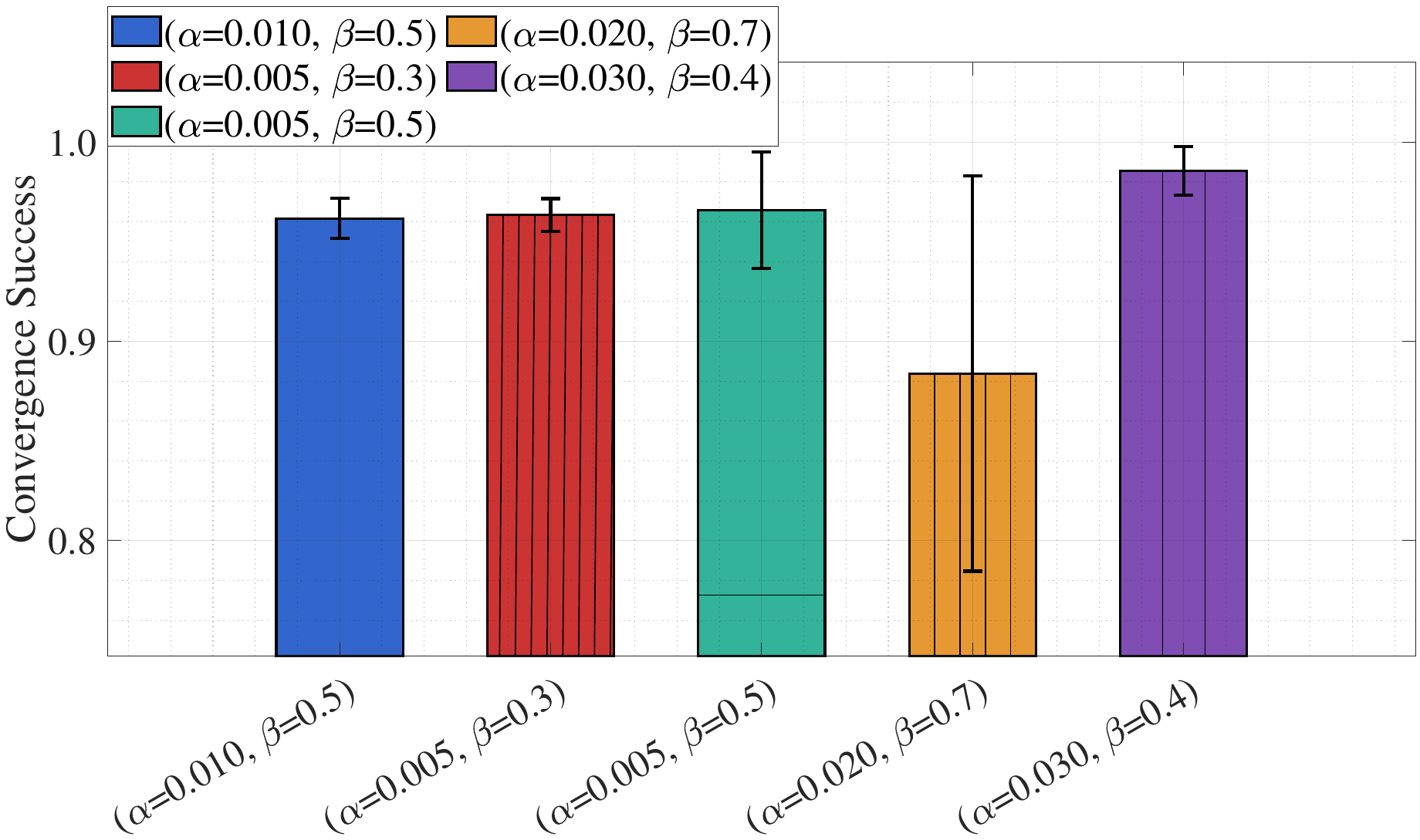}
\caption{Sensitivity Analysis of Regularization Parameters $\alpha$ and $\beta$ for Communication Efficiency in IC3Net.}  
    \label{fig:Sensitivity analysis}
    \vspace{-.5cm}
\end{figure}

Fig. \ref{fig:Sensitivity analysis} presents a sensitivity analysis of the regularization parameters $\alpha$ and $\beta$ in Algorithm 1, using IC3Net as the baseline. In the proposed communication efficiency regularization framework, parameter $\alpha$ controls the relative weight of the IEI regularization term in the total loss, determining how strongly the optimization prioritizes communication efficiency. Parameter $\beta$ scales the success rate in the IEI formulation ($\Phi_{\text{IEI}_t} = H_t \cdot (1.0 - \beta \cdot \mathscr{S}_t)$), modulating the trade-off between message entropy reduction and task performance maintenance. Together, these parameters govern the balance between achieving high coordination performance and improving communication efficiency.

The sensitivity analysis reveals a key finding: despite variations in convergence success rates and training stability across different $(\alpha, \beta)$ configurations, all evaluated parameter pairs yield performance improvements compared to the baseline IC3Net without IEI-based loss adjustment. This demonstrates the robustness of the proposed regularization approach in enhancing communication efficiency while preserving or improving task performance across a range of hyperparameter settings.

Examining specific configurations, $(\alpha=0.005, \beta=0.3)$ achieves high median convergence success rate with the smallest variance, demonstrating excellent balance between performance and training stability. Configuration $(\alpha=0.01, \beta=0.5)$ attains comparable performance with low variance, indicating that moderate regularization strength also maintains stable training. Configuration $(\alpha=0.005, \beta=0.5)$ reaches strong performance but exhibits substantially increased variance, suggesting that higher $\beta$ values with weak $\alpha$ reduce training stability despite achieving good mean performance. Notably, $(\alpha=0.03, \beta=0.4)$ achieves the highest success rate with relatively small variance, demonstrating that stronger $\alpha$ regularization can be effective when paired with moderate $\beta$ values. In contrast, $(\alpha=0.02, \beta=0.7)$ shows significantly degraded performance with the highest variance, indicating that the combination of moderate $\alpha$ with high $\beta$ destabilizes training by over-penalizing necessary communication.

These results reveal that effective communication efficiency regularization requires careful parameter balancing rather than simply moderate settings. While configurations with $\alpha \in [0.005, 0.01]$ and $\beta \in [0.3, 0.5]$ provide stable performance, stronger regularization ($\alpha=0.03$) paired with moderate $\beta$ can achieve superior results, suggesting that the optimal parameter space is more nuanced than initially anticipated.

The IEI regularization mechanism accelerates policy optimization through two complementary effects. First, it constrains the protocol search space by penalizing diffuse message distributions, guiding agents toward compact representations that concentrate information in fewer dimensions~\citep{wang2020learning}, thereby reducing the space of learnable protocols and enabling faster convergence. Second, concentrated message representations yield more focused gradients during backpropagation, channeling optimization pressure toward task-relevant features rather than distributing signals across all dimensions, which reduces gradient variance and stabilizes learning. 
\vspace{-.3cm}
\section{Conclusions}
\vspace{-.3cm}
\label{sec:conclusion}
In this paper, we addressed a key limitation of existing MARL approaches: prioritizing task performance through complex architectures or increased communication overhead without systematically evaluating communication efficiency. We proposed the IEI, a metric quantifying the ratio between message entropy and task performance, enabling agents to learn compact and meaningful communication representations. Our experiments demonstrate that integrating IEI into policy updates allows agents to jointly optimize efficiency and performance, improving convergence speed across diverse MARL algorithms. These findings challenge the assumption that performance improvements require increasingly complex architectures or expanded communication, demonstrating that both objectives can be optimized simultaneously. Our framework provides valuable tools for developing communication-efficient protocols in resource-constrained environments, with future work exploring generalizability across diverse architectures and scenarios where communication constraints critically determine system viability.


\beginSupplementaryMaterials

\noindent\textbf{A. Network Architectures and Hyperparameters}

We evaluate five multi-agent communication algorithms: MAGIC \citep{niu2021multi}, CommNet \citep{sukhbaatar2016learning}, TarMAC \citep{das2019tarmac}, GA-Comm \citep{liu2020multi}, and IC3Net \citep{singh2018learning}. Table~\ref{tab:network_architectures} presents the detailed network architectures for each algorithm, while Table~\ref{tab:training_hyperparameters} summarizes the training hyperparameters.

\begin{table}[htbp]
\centering
\caption{Network Architecture Details for All Baseline Algorithms}
\label{tab:network_architectures}
\small
\begin{tabular}{|l|l|p{7.5cm}|}
\hline
\textbf{Algorithm} & \textbf{Component} & \textbf{Architecture Details} \\
\hline
\hline
\multirow{6}{*}{\textbf{MAGIC}} 
& Observation Encoder & Linear(137 $\rightarrow$ 128) \\
\cline{2-3}
& Recurrent Module & LSTMCell(128, 128) \\
\cline{2-3}
& Graph Attention & GraphAttention(128 $\rightarrow$ 128) \\
\cline{2-3}
& Message Decoder & Linear(128 $\rightarrow$ 128) \\
\cline{2-3}
& Action Head & Linear(256 $\rightarrow$ 2) \\
\cline{2-3}
& Value Head & Linear(256 $\rightarrow$ 1) \\
\hline
\multirow{6}{*}{\textbf{CommNet}} 
& Observation Encoder & Linear(137 $\rightarrow$ 128) \\
\cline{2-3}
& Hidden Encoder & Linear(128 $\rightarrow$ 128) \\
\cline{2-3}
& Recurrent Module & LSTMCell(128, 128) \\
\cline{2-3}
& Communication Module & Linear(128 $\rightarrow$ 128) \\
\cline{2-3}
& Action Head & Linear(128 $\rightarrow$ 2) \\
\cline{2-3}
& Value Head & Linear(128 $\rightarrow$ 1) \\
\hline
\multirow{9}{*}{\textbf{TarMAC}} 
& Observation Encoder & Linear(137 $\rightarrow$ 128) \\
\cline{2-3}
& Hidden Encoder & Linear(128 $\rightarrow$ 128) \\
\cline{2-3}
& Recurrent Module & LSTMCell(128, 128) \\
\cline{2-3}
& Query Network & Linear(128 $\rightarrow$ 16) \\
\cline{2-3}
& Key Network & Linear(128 $\rightarrow$ 16) \\
\cline{2-3}
& Value Network & Linear(128 $\rightarrow$ 128) \\
\cline{2-3}
& Communication Module & Linear(128 $\rightarrow$ 128) \\
\cline{2-3}
& Action Head & Linear(128 $\rightarrow$ 2) $\times$ 2 heads \\
\cline{2-3}
& Value Head & Linear(128 $\rightarrow$ 1) \\
\hline
\multirow{10}{*}{\textbf{GA-Comm}} 
& Observation Encoder & Linear(137 $\rightarrow$ 128) \\
\cline{2-3}
& Hidden Encoder & Linear(128 $\rightarrow$ 128) \\
\cline{2-3}
& Recurrent Module & LSTMCell(128, 128) \\
\cline{2-3}
& Query Network & Linear(128 $\rightarrow$ 16) \\
\cline{2-3}
& Key Network & Linear(128 $\rightarrow$ 16) \\
\cline{2-3}
& Bi-LSTM & LSTM(256 $\rightarrow$ 256, bidirectional) \\
\cline{2-3}
& Gating Network & Linear(512 $\rightarrow$ 2) \\
\cline{2-3}
& Communication Module & Linear(128 $\rightarrow$ 128) \\
\cline{2-3}
& Action Head & Linear(256 $\rightarrow$ 2) \\
\cline{2-3}
& Value Head & Linear(256 $\rightarrow$ 1) \\
\hline
\multirow{6}{*}{\textbf{IC3Net}} 
& Observation Encoder & Linear(137 $\rightarrow$ 128) \\
\cline{2-3}
& Hidden Encoder & Linear(128 $\rightarrow$ 128) \\
\cline{2-3}
& Recurrent Module & LSTMCell(128, 128) \\
\cline{2-3}
& Communication Module & Linear(128 $\rightarrow$ 128) with gating \\
\cline{2-3}
& Action Head & Linear(128 $\rightarrow$ 2) $\times$ 2 heads \\
\cline{2-3}
& Value Head & Linear(128 $\rightarrow$ 1) \\
\hline
\end{tabular}
\end{table}

\begin{table}[htbp]
\centering
\caption{Training Hyperparameters for All Algorithms}
\label{tab:training_hyperparameters}
\small
\begin{tabular}{|l|c|p{6.5cm}|}
\hline
\textbf{Hyperparameter} & \textbf{Value} & \textbf{Description} \\
\hline
\hline
\multicolumn{3}{|c|}{\textit{\textbf{General Training Parameters}}} \\
\hline
Value Loss Coefficient & 0.01 & Weight for value function loss \\
\hline
Gradient Detach Gap & 10 & Steps between gradient detachment \\
\hline
Random Seed & 0, 1, 2 & Seeds for multiple runs \\
\hline
\hline
\multicolumn{3}{|c|}{\textit{\textbf{Network Architecture Parameters}}} \\
\hline
Hidden Size & 128 & Dimension of hidden states \\
\hline
Query/Key Size & 16 & Dimension for attention mechanisms \\
\hline
Value Size & 32 & Dimension for attention values \\
\hline
GAT Attention Heads & 4 & Number of attention heads (MAGIC) \\
\hline
GAT Hidden Size & 32 & Hidden size for GAT layers \\
\hline
Communication Rounds & 1 or 2 & Number of message passing rounds \\
\hline
Communication Init & Uniform & Message initialization strategy \\
\hline
RNN Type & LSTM & Recurrent cell type \\
\hline
\end{tabular}
\end{table}

\vspace{0.5cm}
\noindent\textbf{B. Environment Configuration}

The Traffic Junction (TJ) environment \citep{singh2018learning} simulates a cooperative navigation scenario where multiple agents must cross an intersection without collisions. 

\textbf{Observation Space:} Each agent observes its local neighborhood within a vision range of 1. The observation includes: (1) agent's own position and velocity (2D coordinates), (2) positions and velocities of visible neighboring agents, (3) road structure information (intersection layout), (4) traffic light states (if applicable), and (5) goal direction indicator. The raw observations are concatenated into a 137-dimensional vector.

\textbf{Action Space:} Each agent selects from 2 discrete actions at each timestep: \texttt{Accelerate} (move forward along the current lane) and \texttt{Brake} (stop or slow down to avoid collision).

\textbf{Reward Structure:} The environment provides sparse rewards: $-1$ penalty for each collision between agents, and $0$ reward for successful crossing without collision. Episodes terminate when all agents cross or reach maximum steps.

\vspace{0.5cm}
\noindent\textbf{C. Experiment Implementation Details}

\textbf{Parameter Initialization Strategy:} All network parameters are initialized using the following strategies: (1) \textbf{Linear Layers} use Xavier uniform initialization with gain=1.0; (2) \textbf{LSTM Cells} use orthogonal initialization for recurrent weights and zero initialization for biases; (3) \textbf{Communication Messages} are uniformly randomly initialized in $[-0.1, 0.1]$; (4) \textbf{Attention Weights} use Xavier uniform initialization with gain=$1/\sqrt{d_k}$ where $d_k$ is the key dimension; (5) \textbf{Optimizer} is Adam with $\beta_1=0.9$, $\beta_2=0.999$, $\epsilon=10^{-8}$.

\textbf{Convergence Criteria:} We define convergence based on the following criteria: (1) \textit{Success Rate Stabilization} -- the moving average (window=50 epochs) of success rate changes by less than 0.01 over 100 consecutive epochs; (2) \textit{IEI Stabilization} -- for our proposed method, we additionally require that $\Phi_{\text{IEI}}$ stabilizes (change $< 0.5$ over 100 epochs); (3) \textit{Maximum Epochs} -- training terminates at 2000 epochs regardless of convergence status. We report the best performance achieved during training rather than only the final epoch performance.

\noindent\textbf{D. Communication Efficiency Analysis}

\textbf{D.1. Total Communication Number per Epoch During Training}

To comprehensively evaluate the impact of IEI regularization on communication efficiency, we analyze the evolution of total communication volume throughout the training process. The \textit{Total Communication Number per Epoch} metric quantifies the cumulative number of messages exchanged among agents during each training epoch, providing insights into how the IEI mechanism improves communication efficiency while maintaining or improving performance.
\begin{figure}[h]
  \centering
  \includegraphics[width=1\textwidth]{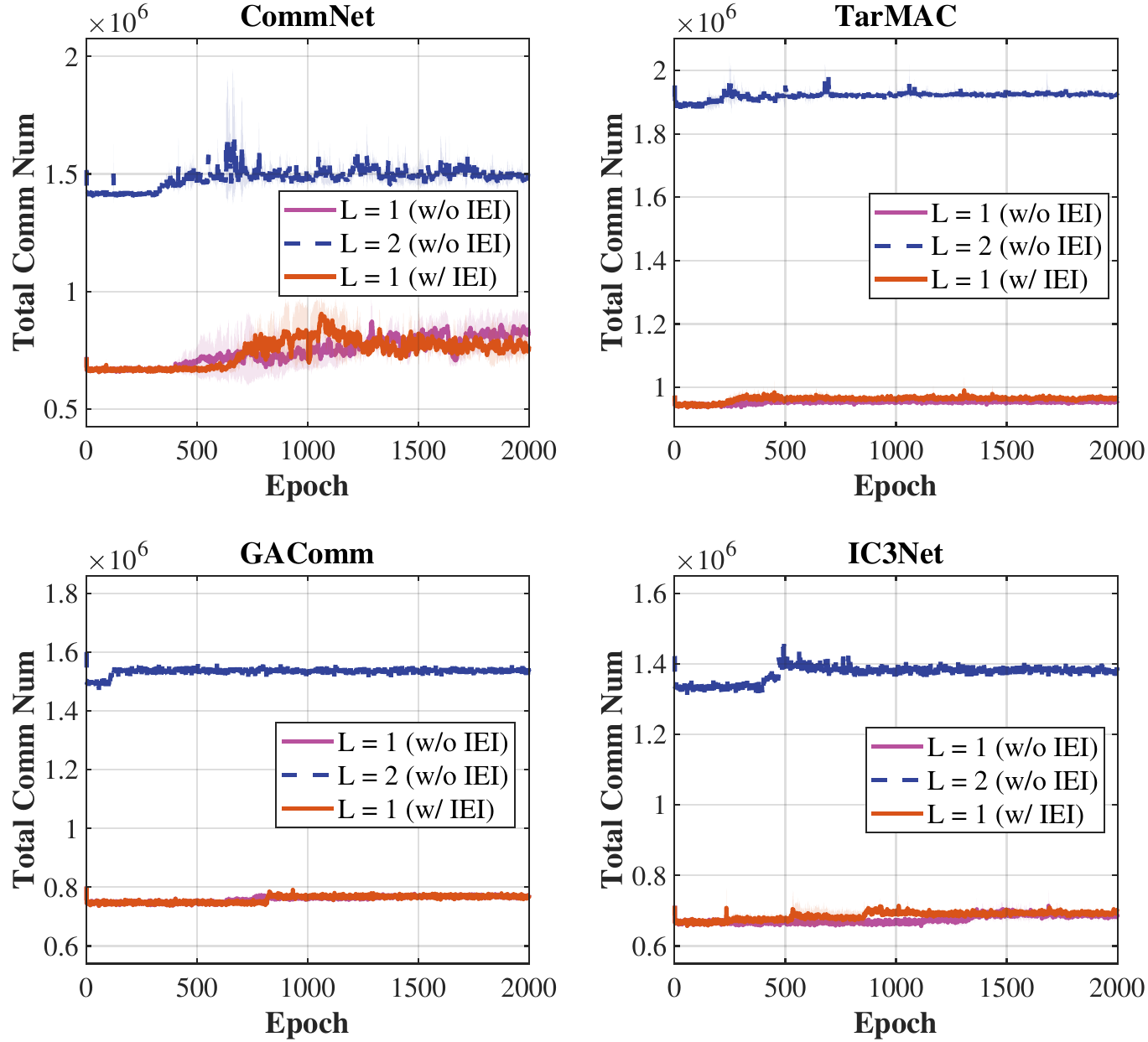}
  \caption{Total Communication Number per Epoch over Three Cases across Different Baselines.}
  \label{fig:Total_comm_num}
  \vspace{-.3cm}
\end{figure}
Figure~\ref{fig:Total_comm_num} illustrates the evolution of total communication volume across training epochs for three different cases. A notable observation is that L=1 (w/o IEI) and L=1 (w/ IEI) maintain comparable communication costs throughout training, while L=2 (w/o IEI) consistently exhibits significantly higher communication overhead due to the additional communication round. This demonstrates that IEI achieves enhanced performance without incurring additional communication burden compared to the baseline single-round approach.

Table~\ref{tab:three_case_comparison} provides a comprehensive quantitative comparison of the five baseline algorithms across multiple performance dimensions. Several key findings emerge from this analysis:

\begin{table}[htbp]
\centering
\caption{Comprehensive Comparison: One Round w/o IEI vs Two Rounds w/o IEI vs One Round w/ IEI}
\label{tab:three_case_comparison}
\small
\begin{tabular}{|c|l|c|c|c|c|}
\hline
\textbf{Algorithm} & \textbf{Case} & \textbf{Max Success} & \textbf{Conv. Epoch} & \textbf{Min Comm} & \textbf{Final Comm} \\
\hline
\hline
\multirow{3}{*}{MAGIC} & L = 1 (w/o IEI) & 0.9044 $\pm$ {\footnotesize 0.0986} & 492 & 737293 $\pm$ {\footnotesize 735} & 832487 $\pm$ {\footnotesize 119891} \\
 & L = 2 (w/o IEI) & 1.0000 $\pm$ {\footnotesize 0.0000} & 1144 & 737637 $\pm$ {\footnotesize 635} & 1537315 $\pm$ {\footnotesize 7230} \\
 & L = 1 (w/ IEI) & 1.0000 $\pm$ {\footnotesize 0.0000} & 474 & 738426 $\pm$ {\footnotesize 0} & 771422 $\pm$ {\footnotesize 0} \\
\hline
\multirow{3}{*}{CommNet} & L = 1 (w/o IEI) & 0.7482 $\pm$ {\footnotesize 0.2920} & 969 & 657197 $\pm$ {\footnotesize 511} & 821304 $\pm$ {\footnotesize 90835} \\
 & L = 2 (w/o IEI) & 0.9991 $\pm$ {\footnotesize 0.0001} & 367 & 1392510 $\pm$ {\footnotesize 2990} & 1477974 $\pm$ {\footnotesize 14015} \\
 & L = 1 (w/ IEI) & 0.9157 $\pm$ {\footnotesize 0.1122} & 1067 & 657117 $\pm$ {\footnotesize 431} & 765237 $\pm$ {\footnotesize 46973} \\
\hline
\multirow{3}{*}{TarMAC} & L = 1 (w/o IEI) & 0.7887 $\pm$ {\footnotesize 0.0028} & 416 & 937246 $\pm$ {\footnotesize 721} & 957932 $\pm$ {\footnotesize 2580} \\
 & L = 2 (w/o IEI) & 0.9945 $\pm$ {\footnotesize 0.0018} & 501 & 1872744 $\pm$ {\footnotesize 634} & 1920432 $\pm$ {\footnotesize 6585} \\
 & L = 1 (w/ IEI) & 0.9828 $\pm$ {\footnotesize 0.0186} & 388 & 936721 $\pm$ {\footnotesize 859} & 965363 $\pm$ {\footnotesize 7049} \\
\hline
\multirow{3}{*}{GAComm} & L = 1 (w/o IEI) & 0.9748 $\pm$ {\footnotesize 0.0285} & 1294 & 736901 $\pm$ {\footnotesize 148} & 773801 $\pm$ {\footnotesize 522} \\
 & L = 2 (w/o IEI) & 0.9610 $\pm$ {\footnotesize 0.0002} & 123 & 1471516 $\pm$ {\footnotesize 0} & 1536860 $\pm$ {\footnotesize 0} \\
 & L = 1 (w/ IEI) & 0.9615 $\pm$ {\footnotesize 0.0001} & 828 & 736876 $\pm$ {\footnotesize 0} & 772184 $\pm$ {\footnotesize 0} \\
\hline
\multirow{3}{*}{IC3Net} & L = 1 (w/o IEI) & 0.8741 $\pm$ {\footnotesize 0.1696} & 1345 & 656786 $\pm$ {\footnotesize 2019} & 682442 $\pm$ {\footnotesize 12360} \\
 & L = 2 (w/o IEI) & 0.9998 $\pm$ {\footnotesize 0.0004} & 472 & 1312500 $\pm$ {\footnotesize 0} & 1382836 $\pm$ {\footnotesize 0} \\
 & L = 1 (w/ IEI) & 0.9975 $\pm$ {\footnotesize 0.0010} & 868 & 657157 $\pm$ {\footnotesize 841} & 699102 $\pm$ {\footnotesize 5504} \\
\hline
\end{tabular}
\end{table}
\textbf{Communication Cost Efficiency:} The results reveal that L=1 (w/ IEI) achieves communication costs comparable to L=1 (w/o IEI) while delivering performance levels approaching or matching L=2 (w/o IEI). For instance, MAGIC with IEI attains 100\% success rate with only 771K messages (Final Comm), compared to 1.54M messages required by L=2 (w/o IEI)-representing a 49.8\% reduction in communication overhead. Similarly, IC3Net with IEI achieves 99.75\% success with 699K messages versus 1.38M messages for L=2 (w/o IEI), yielding a 49.4\% cost saving. Notably, the minimum communication costs (Min Comm) across all three cases remain nearly identical for each algorithm (e.g., MAGIC: 737K-738K, IC3Net: 657K), indicating that the communication cost differences emerge primarily during the learning process rather than being inherent to the communication architecture.

\textbf{Communication Cost Evolution:} The disparity between Min Comm and Final Comm provides insights into communication overhead accumulation during training. L=2 (w/o IEI) consistently exhibits the largest gap between minimum and final communication costs—MAGIC shows an 108\% increase (738K → 1.54M), CommNet 6.1\% (1.39M → 1.48M), and TarMAC 2.5\% (1.87M → 1.92M). In contrast, L=1 (w/ IEI) maintains remarkably stable communication costs throughout training, with MAGIC demonstrating only a 4.5\% increase (738K → 771K) and IC3Net 6.4\% (657K → 699K). This stability suggests that IEI regularization not only reduces absolute communication volume but also promotes more consistent communication patterns across training epochs, potentially indicating more efficient gradient-based learning dynamics.

\textbf{Performance Enhancement:} IEI consistently improves maximum success rates compared to the baseline L=1 (w/o IEI) across all algorithms. MAGIC and IC3Net demonstrate the most substantial improvements, increasing from 90.44\% to 100\% and from 87.41\% to 99.75\%, respectively. TarMAC also shows notable gains, improving from 78.87\% to 98.28\%. CommNet exhibits moderate improvement (74.82\% → 91.57\%), while GAComm maintains stable high performance across all cases (>96\%). These results validate that IEI effectively enhances inter-agent information exchange within a single communication round.

\textbf{Convergence Dynamics:} The convergence epoch metric reveals varied training dynamics across algorithms. MAGIC with IEI converges fastest (474 epochs) while achieving optimal performance. Notably, L=2 (w/o IEI) does not consistently accelerate convergence despite higher communication costs-GAComm converges in only 123 epochs for L=2 but requires 828 epochs with IEI, suggesting that convergence speed is algorithm-dependent and not solely determined by communication frequency. The relationship between convergence speed and final communication cost is non-trivial: while GAComm with L=2 converges rapidly, its final communication cost (1.54M) is nearly double that of L=1 configurations, indicating that faster convergence does not necessarily imply communication efficiency.

\textbf{Training Stability:} The standard deviations in both success rates and communication costs indicate that IEI generally maintains or improves training stability. MAGIC and IC3Net with IEI achieve near-zero variance in final communication costs (±0 for both algorithms) and highly consistent success rates, demonstrating robust and reproducible learning outcomes. The stability in communication costs is particularly noteworthy-L=1 (w/ IEI) configurations exhibit significantly lower variance in Final Comm compared to their L=1 (w/o IEI) counterparts (e.g., CommNet: ±46,973 vs ±90,835; IC3Net: ±5,504 vs ±12,360). In contrast, CommNet exhibits substantial variance across all three cases (e.g., 0.7482 ± 0.2920 for L=1 w/o IEI), suggesting inherent architectural sensitivity to initialization and stochasticity independent of the communication strategy employed.

\textbf{D.2. Communication Efficiency and Pareto Optimality Analysis}

\begin{figure}[h]
  \centering
  \includegraphics[width=1\textwidth]{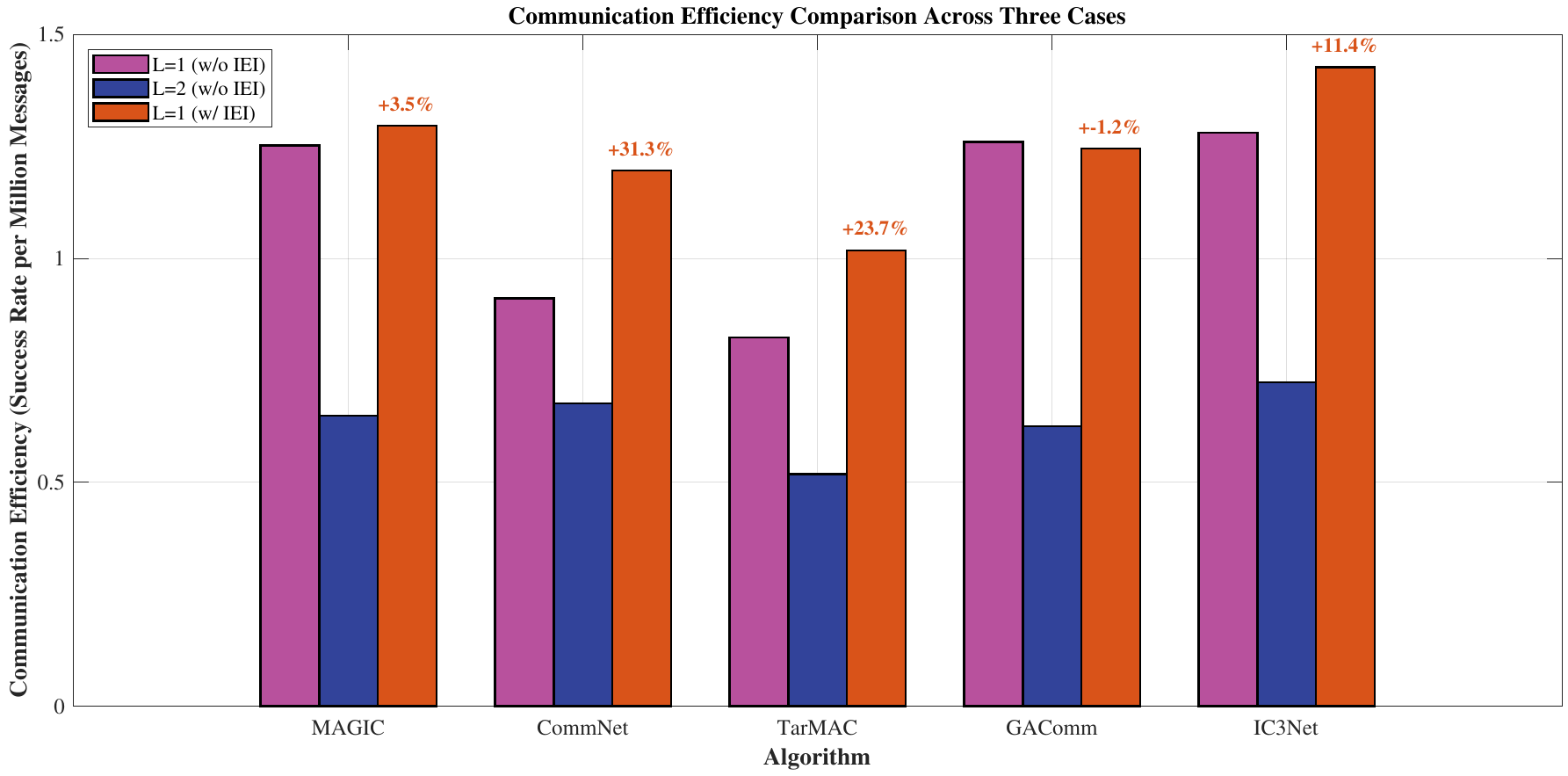}
  \caption{Communication Efficiency Comparison Across Three Cases.}
  \label{fig:Communication Efficiency Comparison}
  \vspace{-.3cm}
\end{figure}

\begin{figure}[h]
  \centering
  \includegraphics[width=1\textwidth]{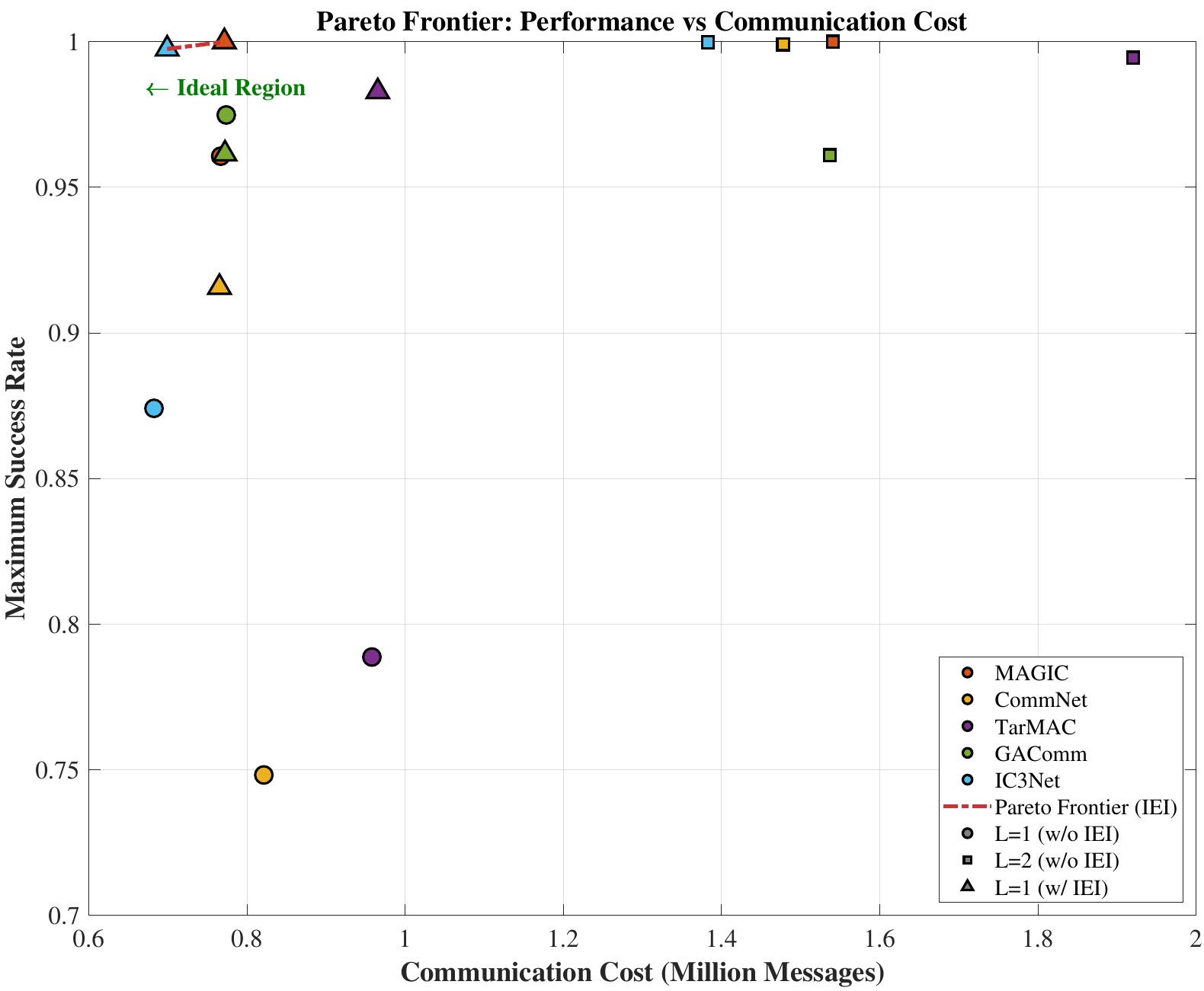}
  \caption{Pareto Frontier: Performance vs Communication Cost across L = 1 (w/o IEI), L = 2 (w/o IEI), and L = 1 (w/ IEI).}
  \label{fig:Pareto Frontier}
  \vspace{-.3cm}
\end{figure}

Figure~\ref{fig:Communication Efficiency Comparison} presents a comparative analysis of communication efficiency across the three experimental cases. Communication efficiency is defined as the ratio of maximum success rate to communication cost (measured in million messages), representing the performance achieved per unit of communication overhead. The results demonstrate that L=1 (w/ IEI) consistently achieves the highest communication efficiency across all five baseline algorithms, with improvements ranging from +3.5\% (MAGIC) to +33.8\% (CommNet) compared to L=1 (w/o IEI). Notably, L=2 (w/o IEI) exhibits substantially lower efficiency due to its doubled communication volume, despite achieving competitive success rates. This finding underscores the fundamental advantage of IEI: enhancing information exchange quality within a single communication round rather than relying on increased communication frequency.

Figure~\ref{fig:Pareto Frontier} further elucidates the performance-cost trade-offs across different configurations through Pareto frontier analysis. Each algorithm is represented by three data points corresponding to the three cases, with marker shapes distinguishing the configurations: circles for L=1 (w/o IEI), squares for L=2 (w/o IEI), and triangles for L=1 (w/ IEI). The red dashed line delineates the Pareto frontier formed by IEI-enabled configurations, indicating the optimal trade-off boundary. Several critical observations emerge: (1) IEI-enabled configurations (triangles) predominantly occupy the ideal region (upper-left), achieving high success rates with minimal communication costs; (2) L=2 configurations (squares) are consistently positioned toward the right side of the plot, reflecting their elevated communication overhead approximately double that of single-round approaches; (3) MAGIC, TarMAC, and IC3Net with IEI achieve near-perfect success rates (>98\%) while maintaining communication costs comparable to their L=1 baseline counterparts, demonstrating superior Pareto optimality.

In summary, the integrated analysis of communication efficiency (Figure~\ref{fig:Communication Efficiency Comparison}), Pareto optimality (Figure~\ref{fig:Pareto Frontier}), and comprehensive metrics (Table~\ref{tab:three_case_comparison}) establishes that IEI achieves a favorable balance between performance and communication efficiency. By enhancing information exchange quality within a single communication round, IEI positions itself as a Pareto-optimal solution that obviates the need for multiple communication rounds while delivering superior task performance with high training stability.

\end{document}